\documentclass[12pt,twoside]{article}
\usepackage[dvips]{graphicx}
\usepackage{cite}

\def\hbabar{\mbox{{\huge\bf\sl B}\hspace{-0.1em}{\LARGE\bf\sl A}\hspace{-0.03em}{\huge\bf\sl B}\hspace{-0.1em}{\LARGE\bf\sl A\hspace{-0.03em}R}}}
\def\lbabar{\mbox{{\large\sl B}\hspace{-0.4em} {\normalsize\sl A}\hspace{-0.03em}{\large\sl B}\hspace{-0.4em} {\normalsize\sl A\hspace{-0.02em}R}}}
\usepackage{relsize}
\def\babar{\mbox{\slshape B\kern-0.1em{\smaller A}\kern-0.1em
    B\kern-0.1em{\smaller A\kern-0.2em R}}}

\newcommand{\epem}    {\ensuremath{\mathrm{e^+e^-}}}
\newcommand{\tmpl}[1] {\mbox{\ensuremath{<\mathrm{#1}>}}}
\newcommand{\btmpl}[1] {\boldmath\tmpl{#1}}
\newcommand{\kanga}   {K{\footnotesize ANGA}}

\begin{document}

%
%  Title Page:
\begin{titlepage}

\pagestyle{empty}

\begin{flushright}
SLAC--PUB--9253\\
BABAR--PUB--02/005\\
June 2002
\end{flushright}

\bigskip
\begin{center}
{\huge\bf K{\Large ANGA}(R{\Large OO}): Handling the \\[2mm]
          micro-DST of the \hbabar\ Experiment \\[3mm]
          with ROOT}
\end{center}

\bigskip

\begin{center}
{\large T. J. Adye$^1$, A. Dorigo$^2$, R. Dubitzky$^3$, A. Forti$^4$,
S. J. Gowdy$^{5,10}$, G. Hamel de Monchenault$^{5,8}$, R. G. Jacobsen$^5$,
D. Kirkby$^{6,11}$, S. Kluth$^{5,9}$, E. Leonardi$^7$, A. Salnikov$^{5,10}$,
L. Wilden$^3$ \\[5mm]
for the \lbabar\ Computing Group }

\end{center}

\begin{center}
{\small
$^1$ Rutherford Appleton Lab, Didcot, Oxon., OX11 0QX, United Kingdom \\
$^2$ Universita di Padova, Dipartimento di Fisica and INFN, Sezione di
Padova Via F. Marzolo 8, I-35131 Padova, Italy \\
$^3$ TU Dresden, Institut f\"ur Kern- und Teilchenphysik, D-01062
Dresden, Germany \\
$^4$ Department of Physics and Astronomy, University of Manchester,
Oxford Road, Manchester, M13 9PL, United Kingdom \\
$^5$ Lawrence Berkeley National Lab, Berkeley, CA 94720, USA \\
$^6$ Stanford Linear Accelerator Center, Stanford University, Stanford, CA
94309, USA \\
$^7$ INFN Sezione di Roma, Dipartimento di Fisica G. Marconi, Piazzale
Aldo Moro 2, I-00185 Rome, Italy \\
$^8$ now at CEN, Saclay, DAPNIA/SPP, F-91191 Gif-sur-Yvette Cedex, France \\
$^9$ now at Max-Planck-Institut f\"ur Physik, F\"ohringer Ring 6, D-80805
Munich, Germany \\
$^{10}$ now at Stanford Linear Accelerator Center, Stanford
University, Stanford, CA 94309, USA\\
$^{11}$ now at UC Irvine, Department of Physics and Astronomy, Irvine, CA 92697-4575 USA
}
\end{center}

\begin{abstract}
\noindent 

A system based on ROOT for handling the micro-DST of the \babar\
experiment\footnote[12]{Work supported by
Department of Energy contract  DE--AC03--76SF00515.}
is described. The purpose of the \kanga\ system is to have
micro-DST data available in a format well suited for data distribution
within a world-wide collaboration with many small sites. The design
requirements, implementation and experience in practice after three
years of data taking by the \babar\ experiment are presented. 

%Some features of the \kanga\ system are discussed in comparison with
%the Objectivity/DB based solution adopted as standard by the \babar\
%collaboration.

\end{abstract}

\begin{center}
{\Large Submitted to Computer Physics Communications}
\end{center}

\end{titlepage}

\tableofcontents

\section{ Introduction }

The \babar\ experiment at the PEP II \epem\ collider located at the
Stanford Linear Accelerator Center (SLAC) studies predominantly CP
violation effects in the neutral B meson system~\cite{bbr-01-18}. The
relative production rate of events with CP violating effects is small
because decay channels of the neutral B mesons with branching ratios
${\cal O}(10^{-3})$ or smaller are involved. The physical effects of
CP violation in the neutral B meson system are subtle and require
careful and precise measurements of properties like particle lifetime
and particle 4-momentum.  For these reasons the
collider and the \babar\ experiment have been designed and constructed to
produce large quantities of well measured events~\cite{bbrtdr}. 

The \babar\ detector has approximately 200k channels of digital
information which are read out with a nominal trigger rate of 100 Hz
and a typical event size of 25 kB (RAW data). The output of
reconstruction of physical properties of the events (REC data), two
levels of data reduction (Event Summary Data ESD and Analysis Object
Data AOD), a set of flags and other quantities for fast preselection
of events (TAG data) and simulated events also contribute to the
amount of data. The combination of AOD, TAG and a subset of the
information from the physics generator in simulated events (micro-truth)
are referred to as {\em micro-DST}. The experiment stored about 300~TB
of data in total in the year 2001. The majority of the data is written
on magnetic tapes while ${\cal O}(10)$~TB of mostly micro-DST are kept
on disk for analysis using SLAC based computer systems.

The large number of collaborators (almost 600) and participating
institutions (74 in 9 countries) in the \babar\ collaboration together
with the large data volumes lead to the need to export data from SLAC
to outside institutes. These exports are mostly done with micro-DST
and are intended to spread the load of physics analysis over
several locations.

The \babar\ collaboration has chosen to implement its necessary
software for data handling, reconstruction and data reduction using
the object oriented method with the programming language
C++~\cite{bbrtdr,stroustrup}. The primary data store uses the
commercial object oriented database system
Objectivity/DB~\cite{objy,rd45_99}. The design of the reconstruction
software has the explicit aim of isolating the data storage mechanism
from the central part, referred to as the {\em framework}, and from
the actual reconstruction and data reduction code.

The first operational experience gained with the \babar\ data store from
tests and with the first data in 1999 indicated several problems
caused by technical limitations of
Objectivity/DB~\cite{bbrdb01,bbrdcm01} at that time.  A {\em
federation}, i.e. the data store as it appears to applications, could
not consist of more than 64k individual database files. Since only one
federation could be accessed by an application the size of the
individual database files had to be made large; database files of
10~GB for bulk data and 2~GB for analysis data were used. This in turn
caused data exportation, in particular of collections of sparsely
distributed events, to be an inefficient and cumbersome process. For
these reasons data exportation of micro-DST to outside institutes did
not take enough load off the computing resources at SLAC. The
turn-around time of data analysis at SLAC were considered to be too
long within the collaboration.

To alleviate the situation it was decided to build an alternative
system independent of Objectivity/DB for the storage of the micro-DST
used in most analyses as a short to medium term solution.  The aim was
to simplify data exportation to remote institutes and thus to improve
access to the data by the collaboration as a whole in order to get
physics results quickly.  The system was named \kanga(R{\footnotesize
OO}), an acronym 
for Kind ANd Gentle Analysis (without Relying On Objectivity).  This
paper describes the design and implementation of the \kanga\ system in
section~\ref{sec_design}, reports experience gained in production
in section~\ref{sec_practice} and presents a discussion in
section~\ref{sec_diss}.  The last section~\ref{sec_conc} gives 
a summary and some conclusions.

\section{ Design and Implementation }
\label{sec_design}

\subsection{ The \babar\ Framework }

In the following we give a brief description of the design of \babar\
framework as relevant for data handling. More detail can be found
in~\cite{rd45_99,bbrfmk97,jacobsen97}.  The object oriented design of
the framework is based on the traditional HEP data processing model.
Event data and the output of reconstruction or data reduction
algorithms are handled via a central object referred to as {\em
event}. The event is accessed using a so called {\em interface
dictionary} through which pointers to data items are stored and
retrieved with the help of keys to identify data items. Algorithms are
mapped on {\em modules} which are managed by the framework.  Modules
get access to the event, retrieve and process data, create new data
from the results of processing and put these new data back into the
event. The framework executes the modules in a configurable sequence
for each event.

Modules are constructed as subclasses of a generic module class which
enforces the existence of member functions for job initialisation,
event processing and job finalisation. The framework also defines more
specialised generic classes (as subclasses of the generic module) for
specific tasks like input and output of data from/to permanent storage
(IO modules) and concrete implementations exist for interaction with
Objectivity/DB. So-called {\em filter modules} can cause the framework
to stop the execution of the module sequence for the current event and
start to process the next event. 

Data items are represented by both transient and persistent
classes~\cite{bbrschema00,rd45_99}. Instances of transient classes
(transient objects) only exist in the computers memory and modules get
only access to transient objects via the event. Instances of
persistent classes (persistent objects) may exist in memory as well as
in permanent storage, e.g. in Objectivity/DB. Persistent objects can
be constructed from transient objects and persistent objects can
create a corresponding transient object via a member function. This
behaviour is controlled by base classes, templated by the
corresponding transient class in the case of the persistent classes.
This design completely decouples the data representation in memory as
visible by the modules from the data store. Replacing the actual data
store only involves the creation of new persistent classes and new IO
modules with associated helper classes. The new persistent classes
must work with the same transient classes.

\subsection{ Requirements and Guidelines }

The most important requirements for the \kanga\ system were:
\begin{itemize}
\item access to the same micro-DST as with the \babar\ data store,
\item compatible with framework and existing user analysis code,
\item fast event filtering using TAG data,
\item simple distribution of data to remote institutes.
\end{itemize}

These requirements could be met by implementing an alternative for the
Objectivity/DB based \babar\ data store as outlined above using the ROOT
system~\cite{root1,root2} as a file based data store for the micro-DST
only. In this way user analysis code only had to be relinked using the
new input and output modules. The actual user-written modules for data
analysis remained unchanged.

\subsection{ The ROOT System }

The ROOT system has been chosen, because it directly supports an
object persistence model using C++. The ROOT object persistence model
is file oriented and thus does not have some of the conveniences nor
the associated overheads of a complete object-oriented database like
Objectivity/DB.  The ROOT object persistence model does not consider
transactions and thus does not have to administer a locking
system. Access control is implemented using file access modes provided
by the operating system instead.

The ROOT system contains structures to organise the storage of many
persistent objects in files. The structure is laid out as a tree (ROOT
class TTree) with branches (ROOT class TBranch) to which objects are
associated. Persistent objects are serialised into branches using a
simple index. For efficient permanent storage access each branch has a
buffer in RAM, which is only written to disk when it is full, possibly
after compression. The size of the buffer can be set at run time.  In
the so-called split-mode ROOT writes each individual data member of a
persistent class via its own buffer to allow for efficient access when
the file is opened using the interactive root application.  Event data
are given by the set of persistent objects with the same index in the
various branches.  A ROOT file can contain several trees or a tree may
be spread by branches over several files.

\subsection{ Event Data }

In the \kanga\ system persistent objects are stored in several branches
of a single tree per file. There is one branch per persistent class or
collection of persistent classes. Separate files may be used for TAG
data on one side and AOD and micro-truth data on the other side.

\begin{figure}[!htb]
\begin{center}
\includegraphics[width=0.75\textwidth]{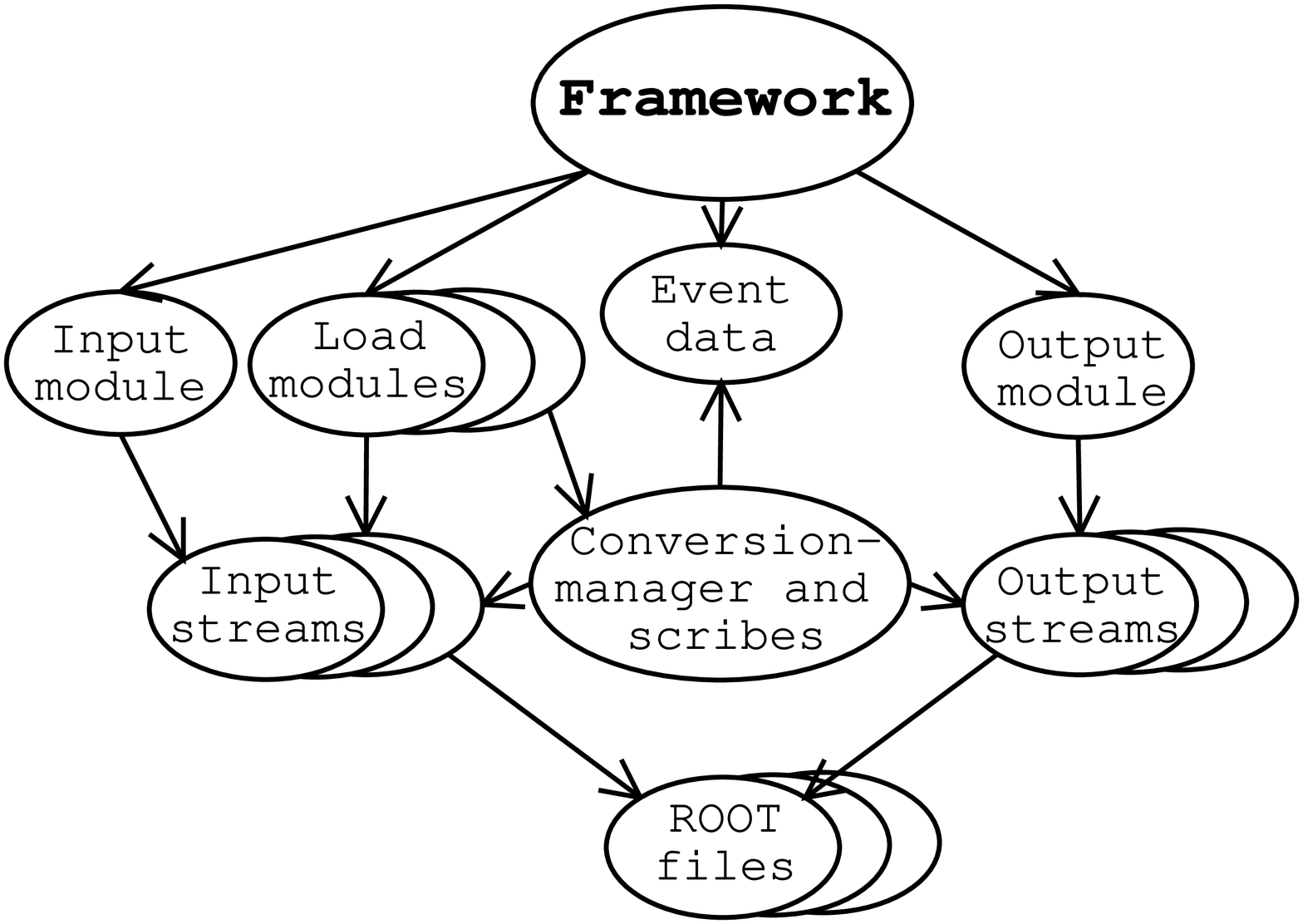}
\caption[ bla ]{ The figure presents an overview of the \kanga\ system
in the \babar\ framework.  The ellipses represent objects or closely
related groups of objects and arrows show how objects use other
objects. }
\label{fig_scenario}
\end{center}
\end{figure}

We provide here an overview of the \kanga\ data handling.  Details of
the implementation are given in appendix~\ref{app_data}.  The handling
of event data in the \kanga\ system follows the implementation of the
\babar\ data store~\cite{rd45_99}.  Figure~\ref{fig_scenario} provides
an overview of the structure. The \babar\ framework controls output and
input modules which in turn control possibly several output and input
streams. The streams handle the actual ROOT files. The event data are
moved between the event object and the streams (and thus ROOT files)
by the conversion manager with the help of scribes. Different
implementations of the scribes handle single objects or collections of
objects in one-to-one, many-to-one or many-to-many mappings between
transient and persistent objects.  Dedicated so-called load modules
for each data type, e.g. TAG or AOD, are responsible for creating the
correct scribes and connecting them with output or input streams.

\begin{figure}[!htb]
\begin{center}
\includegraphics[width=0.75\textwidth]{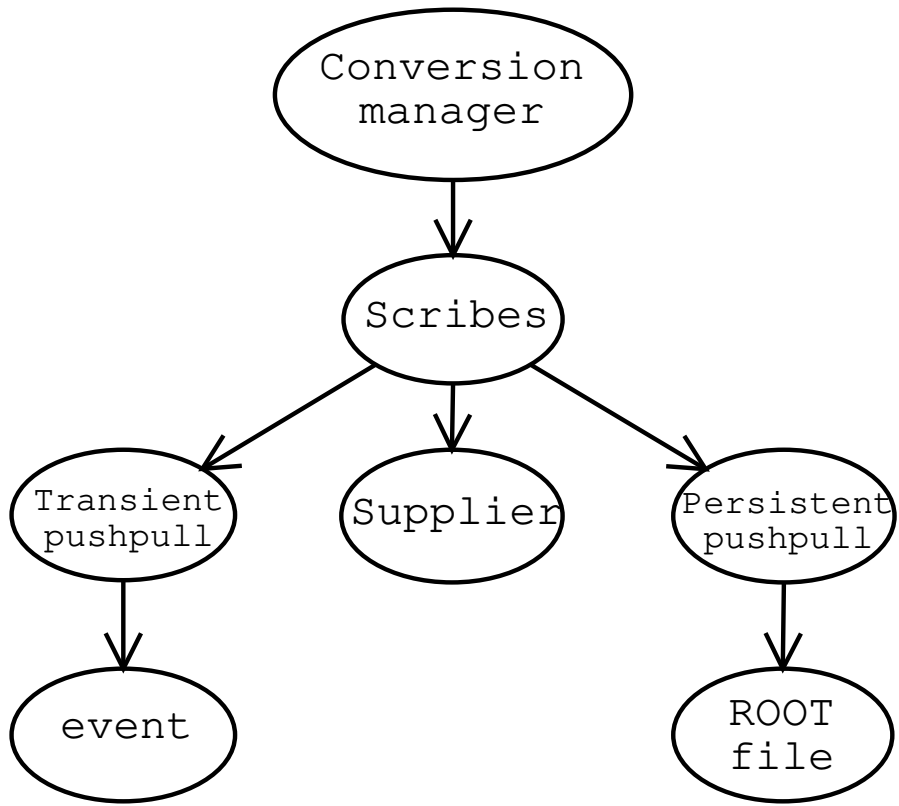}
\caption[ bla ]{ The figure shows an overview of the scribes and
associated objects which handle transient and persistent objects. The
ellipses represent objects and arrows show how objects use other
objects. }
\label{fig_scribes_scen}
\end{center}
\end{figure}

The scribes use helper classes to perform retrieval of transient
objects from the event, retrieval of persistent objects from the ROOT
files and conversion between transient and persistent objects as shown
in Fig.~\ref{fig_scribes_scen}. The transient pushpull object can
move transient objects between the event and the scribe, the supplier 
converts between transient and persistent objects and the persistent
pushpull object moves persistent objects between the scribe and the
\kanga\ data store. 

For collections of static\footnote{By {\em static objects} an
identical and fixed memory allocation for each object of a particular
class is meant.} persistent objects a feature of the ROOT persistency
system is used which allows the creation of the collection 
using already allocated memory when data are read. This can increase
significantly the efficiency of analysis jobs. Details of the
implementation may be found in appendix~\ref{app_data}.

\subsection{ Sequences of Operations for Data Output or Input }

There are three phases in a run of a physics analysis program based on
the \babar\ framework. In the first phase (initialisation) variables are
set and run-time configuration of the job is performed. In the second
phase (event loop) event data are read in and processed. The third
phase (finalisation) deals with final calculations of the analysis,
closing of files and printing of results.

\subsubsection{ Data Output }

In this description we assume that transient objects of the micro-DST
are created by either running the reconstruction and data reduction
modules or by reading the micro-DST from the \babar\ data store.

In the initialisation phase for data output the output module is
configured with named output streams. This results in the creation of
output stream objects controlling trees in ROOT files. Subsequent load
modules are configured with a stream name and names of the branches to
which data will be written. The load modules create scribe objects
matching a given subset of the data, e.g. single objects or
collections of objects, and passed the stream and branch names. The
creation of the scribe objects causes the creation of the appropriate
transient and persistent pushpull and supplier objects.

In the event loop a special module creates a conversion manager object
and stores it in the event. The load modules obtain the conversion
manager from the event and pass it a list of scribes for output. The
output module instructs all its output stream objects to execute their
data output method. On the first event a ROOT file is attached and a
new tree is created. After that the conversion manager is obtained
from the event and the conversion manager's method to convert
transient to persistent objects and store them in the ROOT data store
is run.  In this method all scribes connected to the active output
stream are triggered to run their sequence of operations for storage
of persistent objects. In this sequence the scribe's transient
pushpull object is used to get the appropriate transient object from
the event. The object is given to the supplier to create a persistent
object. The persistent pushpull object stores the persistent object in
the proper branch of the tree in the ROOT file. At the end of the
module processing sequence the conversion manager is deleted causing
all scribes to be reset.

The finalisation phase involves essentially only the closing of the
ROOT files.

\subsubsection{ Data Input }

In this description we assume that ROOT files containing micro-DST
already exist. The ROOT files are accessed in an analysis job directly
using their complete directory path and name. Other tools are used to
obtain lists of file names corresponding to selection criteria like
run numbers, event types or special analysis working group selections,
see section~\ref{sec_dataman} below. 

In the initialisation phase the input module is configured with named
input streams and a list of ROOT file names.  This results in the
creation of input stream objects controlling trees in ROOT files in a
read-only mode. Subsequent load modules are configured with a stream
name and names of the branches from which data will be read. The load
modules create scribe objects in exactly the same way as for data
output.

In the event loop a dedicated module creates a new conversion manager
object and stores it in the event. Then the input module stores a list
of its input stream objects in the event. The load modules get the
conversion manager from the event and pass it lists of their scribes
for data input. After that another module gets the conversion manager
and the list of input stream objects from the event and runs the
conversion manager's method to convert persistent to transient objects
for all input streams. The conversion manager causes the scribes in
its list to perform the sequence of operations for data input. In this
sequence the scribe's persistent pushpull object reads the persistent
object from the \kanga\ data store, the persistent object is passed to
the scribe's supplier to create a new transient object and the
transient object is stored in the event by the scribe's transient
pushpull object. At the end of the module processing sequence the
conversion manager is deleted causing all scribes to be reset.

The finalisation phase involves essentially only the closing of the
ROOT files.

\subsection{ TAG Data }

\begin{figure}[!htb]
\begin{center}
\includegraphics[width=\textwidth]{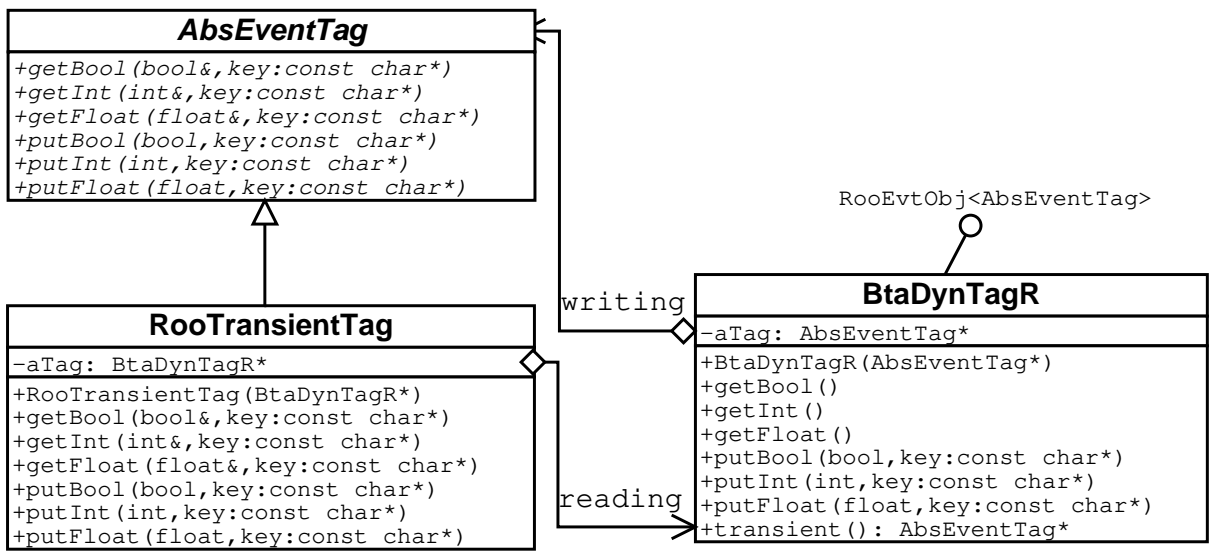}
\caption[ bla ]{ The figure shows an UML class diagram of the handling
of TAG data in the \kanga\ system. }
\label{fig_tag}
\end{center}
\end{figure}

The TAG data are handled in a somewhat different way compared to the
other types of data in the micro-DST in order to accommodate the
special requirements of fast event selection using only a few of many
items in the TAG data. TAG data consist exclusively of boolean,
integer and float types of data items.  The transient class
representing the TAG data is constructed as a so-called
adapter~\cite{gamma} to the persistent TAG data. This leads to a
dependency of the transient TAG data class on the persistent TAG data
class. However, the framework provides an interface for access to the
transient TAG data such that the dependency of the transient TAG data
class on the persistent TAG data class is shielded from the user
modules. The following classes implement the TAG data in
\kanga. Figure~\ref{fig_tag} shows a UML class diagram of the transient
and persistent classes described briefly below.

\begin{description}

\item[BtaDynTagR] This class manages the persistent representation of
TAG data in the tree. It controls one branch for each TAG data item
such that TAG data items can be read individually and it implements
methods which are used by RooTransientTag.

\item[RooTransientTag] This class implements the framework interface
AbsEventTag for access to TAG data. It uses methods of BtaDynTagR to
access TAG data items when e.g. a user module takes an instance of
this class stored in the event to obtain TAG data.

\item[RooTagScribe] This scribe holds a RooDefTransPushPull, a
RooDefSupplier and a Roo-TagPersPushPull object to handle reading and
writing of TAG data. 

\item[RooTagPersPushPull] This specialised pushpull class inherits
from RooDefPersPushPull and implements adapted methods to read (write)
TAG data from (to) the \kanga\ data store (pullPersistent and
pushPersistent).

\end{description}

Writing and reading TAG data uses the same mechanism of scribes,
streams and the conversion manager as for other data. The following
gives a brief description of the sequence of operations concentrating
on points where TAG data are handled differently.

When TAG data are written a dedicated load module creates a
RooTagScribe and passes it to the conversion manager. The RooTagScribe
may be connected to a separate output stream and thus to a separate
tree possibly on a separate ROOT file.  A persistent BtaDynTagR object
is created by the RooTagScribe using the transient AbsEventTag object
from the event.  The persistent object keeps a reference to the
transient object and uses the transients methods to get all TAG data.
A branch is created and filled in the tree for each TAG data item.

When TAG data are read a RooTagScribe created by the load module
and connected to an input stream is used by the
conversion manager. The scribes RooTagPersPushPull object creates a
BtaDynTagR object. A transient RooTransientTag object is created by
the RooDefSupplier using a method of BtaDynTagR and the transient
object is stored in the event. The transient object keeps a reference
to the persistent object from which it was created.  User analysis
modules access the RooTransientTag object in the event through its
interface AbsEventTag and the RooTransientTag object relays the
queries for TAG data items directly to its BtaDynTagR object. The
BtaDynTagR object in turn accesses the branch in the tree containing
the requested TAG data item.

\subsection{ References between Objects }

Transient objects may reference each other, i.e. transient objects may
keep pointers to other transient objects. When transient objects are
converted to persistent objects and stored it is not sufficient to
simply store the values of the pointers, because in the reading job
the memory locations for the new transient objects will be different.
It is thus necessary to have a mechanism to record transient object
relations in the data store and to recreate these relations when
reading data.

\begin{figure}[!htb]
\begin{center}
\includegraphics[width=\textwidth]{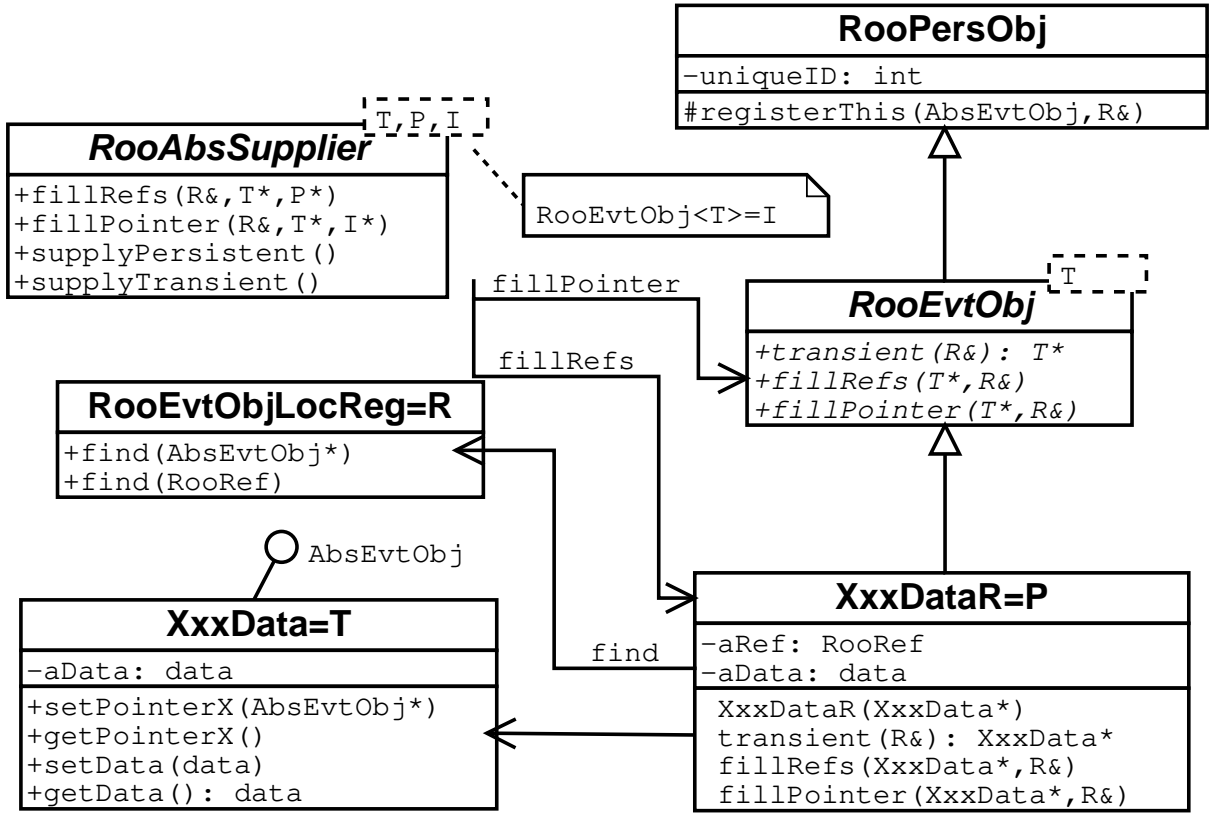}
\caption[ bla ]{ The figure presents an UML class diagram of the
classes involved in the implementation of persistent references.  }
\label{fig_persistent}
\end{center}
\end{figure}

There are two classes to implement the mechanism together with the
classes Roo-PersObj and RooEvtObj\tmpl{T} from which all persistent
classes are derived (see section~\ref{sec_pers}).
Figure~\ref{fig_persistent} shows a UML class diagram.

\begin{description}

\item[RooRef] This class implements a persistent reference as a unique
object identifier. It is used as a data member of a persistent object
to record a relation to another persistent object.

\item[RooEvtObjLocReg] This class is a registry of relations between
transient and persistent objects. It provides bi-directional mapping
between persistent references RooRef and pointers to the transient
objects.

\item[RooPersObj] This class has a method to return its unique object
identifier as a RooRef object and another method to insert a transient
to persistent relation in the registry  to be used in the
constructors of concrete persistent classes. 

\item[RooEvtObj\btmpl{T}] This class provides a method to set
persistent references (of type RooRef) using the transient object T
and the registry RooEvtObjLocReg (fillRefs). It also provides a method
to set transient references in the transient object T from persistent
references and the registry (fillPointers).

\end{description}

In a job which writes data the map of transient-to-persistent object
relations in RooEvtObjLocReg is filled when persistent objects are
constructed from transient objects.  After all persistent objects
are created the conversion manager calls the scribe's method to fill
any persistent references (fillRefs). The task is delegated to the
scribe's supplier which calls the persistent object's method (fillRefs)
to obtain from the transient object any pointers to other
transient objects, map them to the corresponding persistent
references and update the data members of the persistent object using
the resulting RooRef objects. 

When data are read from the \kanga\ data store the map of
persistent-to-transient relations is filled in the methods of the
persistent objects which create new transient objects.  Once all
transient objects are created the conversion manager uses the method
of the scribes to set up references between the transient objects
(fillPointers). The task is delegated by the scribe to its supplier
which in turn uses a method of the persistent object
(fillPointers). In this method any references to other persistent
objects are mapped to the corresponding transient objects and the
resulting pointers to other transient objects are passed to the
transient object created by the persistent object.

This scheme handles references between transient objects valid during
the execution of a job, because the object identification is only
unique within one session.  The scheme is not able to provide unique
object relations within the whole \kanga\ data store since it consists
of many files written by many separate jobs.  The current
implementation of the \kanga\ system does not use this facility,
because there are no relations between transient objects that need
recording in the micro-DST.

\subsection{ Schema Evolution }

The structure of a persistent class may be subject to change,
e.g. because a more space-efficient representation of the data is
introduced or because new data members are added or obsolete ones
are removed. The layout of a class in terms of its data members is
referred to as {\em schema} and changes in the schema are known as
{\em schema evolution}. 

The mechanism for schema evolution is similar to the one used in the
\babar\ data store~\cite{bbrschema00,rd45_99}.  In this scheme explicit
new versions of persistent classes are created when a change in schema
is necessary; already existing persistent classes are not changed.
New versions of a persistent class are required to inherit from the
same interface (Roo-EvtObj\tmpl{T}).  After a change in schema,
i.e. when a new version of a persistent class supersedes an old one,
jobs write only the new versions to the data store.  This is
accomplished by changing the creation of the scribes in the load
module classes.  When data are read, ROOT creates automatically the
correct version of the persistent class using its build-in
RTTI\footnote{Run-Time-Type-Identification} system. The scribes and
the pushpull and supplier classes handle persistent objects only
through their interface when they have been read from the \kanga\ data
store. Thus instances of all versions of a persistent class are
converted to transient objects T through the same interface, usually
RooEvtObj\tmpl{T}.

In this way new programs can read old data in a transparent way as
long as all versions of the persistent class are linked into the
program.  Existing programs, e.g. user analysis jobs, must be relinked
to allow them to read data which has been written with a new schema.

\subsection{ Conditions Data }

Conditions data contain monitored values like high voltages, gas
flows, temperatures, electrical currents etc. describing the state of
the experiment as a function of time.  These data are needed to make
calibrations and corrections which in turn allow the computation of
physics quantities from the RAW data. When micro-DST is produced most
condition information has already been used such that an user analysis
job needs only access to a limited set of conditions data.

In the \babar\ framework events are uniquely identified by a time stamp
with a resolution of about 1~ns. Lookup of conditions data works
in a similar way to the retrieval of event data in the
framework~\cite{bbrcdb00}. A central object called {\em environment}
stores transient objects containing conditions data and is accessed
through an interface dictionary using time stamps as keys. A so called
{\em proxy} object actually handles the transient objects and records
the validity time interval for a given transient object. When a
transient object is requested with a time stamp key outside the
validity time interval the proxy tries to find the persistent object
with the correct validity interval in the conditions database,
converts it to the corresponding transient object and returns it.

\begin{figure}[!htb]
\begin{center}
\includegraphics[width=0.75\textwidth]{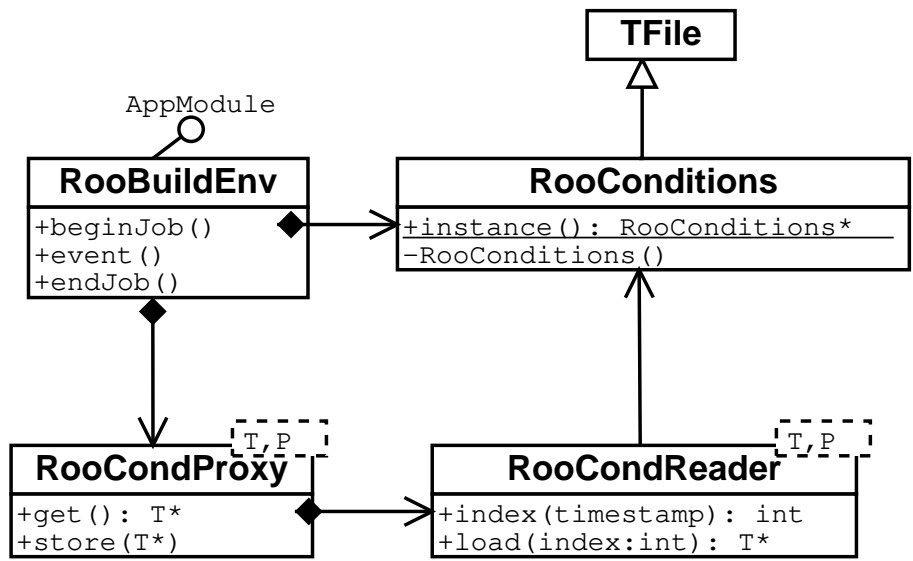}
\caption[ bla ]{ The figure shows an UML class diagram of the
conditions data handling.  }
\label{fig_roocond}
\end{center}
\end{figure}

In the \kanga\ system read-only access to a single revision of the few
conditions data items 
needed by a user analysis job using micro-DST is supported. The ROOT
based conditions database is created by a special application which
reads the latest revision of data from the conditions database in
Objectivity/DB and writes them to a ROOT conditions database file.
In the case that other than the latest revisions of the conditions
data are needed one has to create a new ROOT conditions database
containing the desired revisions of the conditions data. 
The following classes implement the ROOT conditions database, see
Fig.~\ref{fig_roocond} for a UML class diagram.

\begin{description}

\item[RooConditions] This class controls a ROOT file containing the
conditions data. The file contains a tree and an index map for each
type of conditions data. The tree contains in a single branch the
persistent objects in sequential order of insertion. The index map
provides a map of time stamps to indices into the branch of the tree
for time stamp based retrieval of persistent objects. The index map is
a binary tree (ROOT class BTree) of exclusive validity intervals with
a resolution of 1~s. 

\item[RooCondProxy\btmpl{T,P}] This class inherits from the proxy
class of the interface dictionary and provides special implementations
of methods to return transient objects T. The task of retrieving
persistent objects P from the ROOT conditions database is delegated
to RooCondReader\tmpl{T,P} objects.

\item[RooCondReader\btmpl{T,P}] This class controls a tree and its
corresponding index map for persistent objects P. It performs the
translation of a given time stamp in a query to an index in the tree
using the index map, gets the corresponding persistent object P from
the tree, creates a new transient object T from P and returns it.

\item[RooBuildEnv] Instances of this module class create the
RooConditions and the RooCond-Proxy\tmpl{T,P} objects for transient
objects T and corresponding persistent objects P and install them in
the environment in the initialisation method.

\item[RooCondWriter\btmpl{T,P}] This class implements a base class
from the framework which supports writing of conditions data to a new
database. It opens a new ROOT file and creates a tree and an index map
for each type of conditions data to be written. A method from a
framework class then uses methods of this class to write conditions
data to the ROOT conditions database file. 

\end{description}

In the initialisation phase of a job the RooConditions object is
created by the RooBuild-Env module. The RooConditions object will try
to attach a ROOT file with the conditions data. 
 Then RooBuildEnv will create a RooCondProxy\tmpl{T,P}
for each conditions data class T and corresponding persistent class P
and install the proxies in the environment.

In the event loop e.g. a user analysis module may query for a
transient conditions data object with a time stamp as a key.  Such a
query happens when the user module tries to access a transient
conditions data object in the environment.  The key is tested by the
corresponding RooCondProxy against the validity interval of possibly
existing transient objects. When a valid transient object exists
it is returned, otherwise the proxy uses its RooCondReader to obtain
the corresponding persistent object from the ROOT conditions database.
The RooCondReader creates a new transient object which is returned by
the proxy to the user.

In the finalisation phase the ROOT conditions database file is
closed. 

Early versions of \kanga\ involved manual procedures for ensuring
consistency of the conditions and event information.  Physicists found
this sufficiently complicated such that a new way of accessing the
conditions data has been developed.  In this new system
the correct revision of the conditions data is selected
automatically based on information from the collection database (see
section~\ref{sec_dataman} below).

\subsection{ Event Collections }

The \kanga\ system supports the concept of event collections,
i.e. random access to events in root files driven by lists of event
identifiers with pointers to the actual event locations in the ROOT
files.  Event collections are usually created with loose event selection
criteria based on the TAG data with the aim to make repeated analysis
of the preselected events more efficient.  ROOT files with data
organised in trees are well suited for fast random access as long as
the index of the event data in the tree is known. The mapping of
events uniquely identified by their time stamps to indices into the
trees of the ROOT data files is the essential content of an event
collection. Collections of event collections (meta collections) are
transparently implemented by mapping events in a meta collection on
indices in the underlying event collections. In this context trees in
ROOT files containing the event data are also viewed as event
collections. 

\begin{figure}[!htb]
\begin{center}
\includegraphics[width=\textwidth]{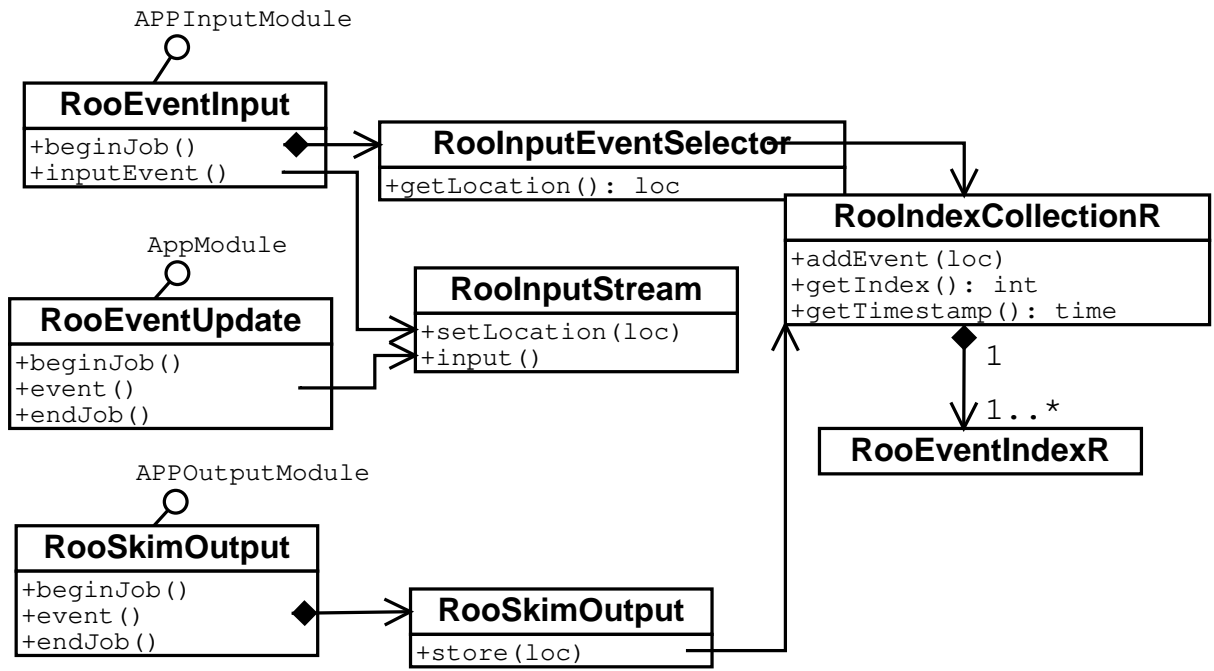}
\caption[ bla ]{ The figure presents an UML class diagram of the
implementation of event collections.  }
\label{fig_collection}
\end{center}
\end{figure}

The following classes are used to implement the writing and reading of
event collections, Fig.~\ref{fig_collection} shows the UML class
diagram.

\begin{description}

\item[RooEventIndexR] This is a class which stores the event time
stamp, an index into a event collection and an index to the name of
the event collection in a ROOT file.

\item[RooIndexCollectionR] This class implements an event
collection. It contains an array of RooEventIndexR and an array of
collection names and can be stored in a ROOT file.  Applications can
iterate over the RooEventIndexR objects and obtain the collection
name, the index into the collection and the event time stamp for each
entry.

\item[RooSkimOutputModule] This is an alternative to the regular
RooOutputModule framework module. It implements the same framework
interface for output modules and delegates the task of creating and
filling of a RooIndexCollectionR to a RooSkim-Output object.

\item[RooSkimOutput] This class opens a ROOT file and controls a
RooIndexCollectionR object. It stores the event time stamp and the
index into the input collection via a RooEventIndexR object in the
RooIndexCollectionR. 

\item[RooInputEventSelector] An instance of this class is used by the
RooEventInput module to find events contained in an event
collection. 

\end{description}

When a collection is written the RooSkimOutputModule is placed
downstream of filter modules in the framework module processing
sequence.  The RooSkimOutputModule creates a RooSkimOutput object on
the first event which comes through. The RooSkimOutputModule opens a
ROOT file and creates a new RooIndexCollectionR to be stored in that
file. After that it obtains the event time stamp, the name of the
input collection and the index of the event in that collection and
stores this information via a RooEventIndexR object in the event
collection.

When an event collection is read, a RooInputEventSelector object is
created by the RooEventInput module in the initialisation phase. This
object is used by RooEventInput in the event loop to locate events
using the information from the event collection. The
RooInputEventSelector descends through any hierarchies of event
collections until the original location of an event as an index into a
tree is found. The event location is passed to all RooInputStreams
controlled by the RooEventInput module, which take care of any
necessary closing and opening of files in the \kanga\ data store.

The current version of \kanga\ does not yet use event collections as
described above in data production. Event collections are realised
more simply as separate ROOT files containing events satisfying
certain selection criteria.

\subsection{ Data Management and Exportation }
\label{sec_dataman}

The mechanism for exportation of data is a central part of the \kanga\
system.  The aim of data exportation is to transfer only the data
files needed at remote sites for processing a given set of event
collections and to replicate the directory structure existing at SLAC 
in the \kanga\ data store only for the selected files. 

A central part of the data exportation system is a relational
database, the {\em collection database}, which is maintained at SLAC
to keep track of data production for the \babar\ data
store~\cite{bbrdb00}.  This database is used in the \kanga\ system to
store additional information about ROOT event data files produced at
SLAC, for example run number, event collection, complete directory path and
file name in the \kanga\ data store, software versions etc.

Tools exist which allow users to query the collection database for
lists of ROOT files corresponding to selection criteria like run
numbers and event collections.  A data analysis thus usually involves
two steps: i) a list of ROOT files present in the \kanga\ data store is
produced by querying the collection database, and ii) this list of
ROOT files is passed as input to the actual framework analysis
program.

Tools for distributing files across a wide area network (WAN) have
been developed, which exploit the information stored in the collection
database at production time~\cite{bbrdex01}.  In the first step of
data exportation the relevant information is extracted from the SLAC
based collection database and transferred to a relational database at
the remote location.  The updated remote collection database is used
to find references to ROOT files which are missing and a list of files
for importation is constructed.  With additional selection criteria
the selection of files for import can be refined.  The files in the
resulting list are transferred over the WAN using a variety of file
transfer tools.  These tools~\cite{bbrdex01} allow to transfer data
using multiple streams and the optimisation of parameters of the WAN
protocol, e.g. the TCP window size.  After a successful update of a
local \kanga\ data store the same tools as at SLAC can be used to
generate lists of input files for analysis jobs using the local \kanga\
data store. 

In early versions of \kanga\ another method based on the rsync
tool~\cite{rsync} was used for data exportation.  The rsync tool
essentially compares two directory trees A and B, determines the
differences between A and B and transfers these differences to
synchronise A with B.  A remote site would synchronise its local
directory tree containing ROOT files with the one at SLAC at regular
intervals.  However, it turned out that with large directory trees
containing ${\cal O}(1000)$ or more files the process of determining
the differences could take several hours.  The method described above
using the collection database scales well to large directory trees. 

The data are organised into {\em skims} and {\em streams}.  A skim
represents a specific event selection while a stream is a physical
output stream corresponding to a ROOT data file.  Several related
skims with many common events are usually grouped into a stream.  In
the year 2001 \babar\ had more than 65 skims grouped into 20 separate
streams while in 2002 more than 100 skims were used.  The
availability of preselected data in streams allows remote sites to
import only the data needed locally for analysis.  There is a
significant number of duplicated events in the several streams leading
to inefficient usage of storage capacity at the SLAC computing centre
and other computing centres keeping a copy of most or all streams in
their \kanga\ data store.

\section{ Practical Experience }
\label{sec_practice}

The \kanga\ system has been in operation in the years 2000, 2001 and
2002.  During the years 2000 and 2001 the \babar\ experiment took a
large amount of data corresponding to 31 million B-meson pairs. During
data taking the micro-DST produced by the online prompt reconstruction
(OPR) in the \babar\ data store was processed offline to generate ROOT
files with a short delay of not more than a few days. Processing was
done on a per-run basis with multiple output streams for different
analysis channels and resulted in ROOT files with a size of a few
100~MB.  From the data of the years 1999, 2000 and 2001 the number of
processed runs was 378, 3333 and 3782, respectively, resulting in
about $3\cdot10^5$ ROOT files occupying about 4.2~TB of space.  The
files are organised in a hierarchical directory structure distributed
over several NFS servers at SLAC.  To balance the load on these servers
when many analysis jobs access the data tools were developed to
distribute the files uniformly on the various disks. The ROOT
conditions database occupies about 20~MB and is kept in a single file.
The copies of the collection database at remote sites needed about
400~MB.

It was verified that user analysis jobs obtain the same results when
processing the same micro-DST from the \babar\ data store or the \kanga\
data store. In the initial commissioning of the \kanga\ system
problems were identified in this way. In particular data are packed in
the \babar\ data store with some tolerable loss in numerical
precision. In initial versions of \kanga\ the data were not packed and
only the lossless compression of ROOT was applied, leading to small
differences in results. \kanga\ now uses the same data packing as in
the Objectivity/DB data store to ensure compatibility of the data
sets and to profit from the additional space savings. 

The average size of the micro-DST per event was initially about
2.3~kB.  The event size varied from about 0.8~kB for a muon-pair event
to about 4~kB for a hadronic event.  The generator information from
the simulation added about 4~kB per hadronic event.  After the
introduction of data packing the average event size was 1.7~kB
corresponding to $21.6~\mathrm{GB}/(fb^{-1})$ of recorded data.  A
representative mix of simulated events occupied 4.7~kB per event.
Production jobs for micro-DST ROOT files ran at SLAC with an event
rate of about $5-20$~Hz depending on the platform and the stream.

The sizes of the internal ROOT buffers connected with each branch were
set to 1024 bytes for TAG branches and to 32768 bytes for other
branches.  These values worked well with serial access to the data,
since data for many events could be read with a single disk access
operation.  Some persistent classes composed mainly of basic types
were written in split-mode to allow for efficient access in an
interactive application.  Using the split-mode was not observed to
cause significant slow-down of analysis jobs.

It was possible to build user analysis jobs without any presence
of Objectivity software or libraries. This allowed remote institutes
to run micro-DST based analysis without the need for an Objectivity
installation. 

Fast event filtering based on the TAG data was implemented using
dedicated event samples based on skims and streams.  A job was run
which read the micro-DST and wrote out the data for events passing TAG
based selection criteria (skims) into streams.  The actual analysis
job was run with the stream containing the skim representing the
desired TAG selection as input.

In total at least 19 institutions operated a local \kanga\ data store.
Out of these 5 had the majority of data available while the other 14
kept only smaller subsamples.  On average 5 skims where imported by a
remote site.  The storage overhead at sites keeping all streams was
about 200 \%, i.e. twice the the size of the sample of all events
considered for physics analysis was needed to make all streams
available due to multiply stored events.  This disadvantage of
inefficient storage was partially offset by the more simple
distribution of special event samples in self-contained files.

In comparison with the standard analysis jobs using \babar\ data store
based micro-DST analysis jobs based on \kanga\ were able to run at
remote institutes on relatively small computer systems compared to the
SLAC computer system.  At SLAC the use of \kanga\ in addition to the
Objectivity/DB based \babar\ data store improved the balancing of
resource loads and therefore made larger overall throughput for
analyses possible.

\section{ Discussion }
\label{sec_diss}

A possible disadvantage of the \kanga\ system is that there is no
direct connection with the ESD or RAW data as is the case with
Objectivity/DB based analyses. However, this restriction can be
circumvented by writing out the event time stamp and reading these
events from the Objectivity/DB \babar\ data store in a separate
application.  Operation of \kanga\ and the \babar\ data store in
parallel required the management of two independent database schemata
and the support of two data storage systems leading to some
duplication of effort.

The system has worked well with the 4.2 TB of micro-DST produced in
the years 1999 to 2001.  In the coming years the \babar\ collaboration
expects a dramatic increase in data volume due to major improvements
in the PEP II collider leading to a much higher luminosity.  The data
sample is expected to grow by about a factor of ten.  With current
technology it seems difficult to keep all micro-DST available on disk
servers as is currently assumed in \kanga.  The \kanga\ system has no
provision for an automatic staging of data from offline storage,
e.g. magnetic tapes.  This limitation could conceivably be lifted by
the deployment of an automatic staging system,
e.g. CASTOR~\cite{castor}.  Files would be registered in CASTOR, which
automatically transfers them to offline storage when necessary.  When
a reference is made to a file registered in CASTOR the system
retrieves the file from the offline storage when it is not present on
disk.  This mechanism would give \kanga\ based analysis jobs transparent
access to ROOT files stored offline and thus allow \kanga\ to operate
with much larger data volumes.

Currently a system to use event collections for skims and streams is
under development.  Only one copy of all events used for analysis
would be kept while the skims and streams would realised as event
collections pointing to events in the master copy.  This would reduce
the necessary disk capacity by about a factor of 2/3.  The system
will still allow to export copies of individual skims or streams
containing the actual event data to remote sites.

\section{ Summary and Conclusions }
\label{sec_conc}

The \babar\ experiment has implemented a data handling system for
micro-DST which is an alternative to the standard system based on
Objectivity/DB. The system was designed to use existing \babar\ software
as much as possible, in particular it was constructed to work within
the \babar\ framework for data reconstruction and analysis programs. The
\kanga\ system uses ROOT as the underlying data storage system, because
ROOT has a simple and efficient object persistency model based on
files.  The \kanga\ system has met its major requirements and was used
successfully in production in the years 2000, 2001 and 2002.

One major aim of the \kanga\ system was the simplification of
exportation of data to remote institutes of the \babar\
collaboration. This aim was reached, because the file oriented data
storage combined with a simple event collection database was easily
adapted to data distribution.

\section*{ Acknowledgments }
The success of the \kanga\ project would not have been possible
without the previous work of the \babar\ data store development team.

\appendix

\section*{ Appendix }

\section{ Event Data Handling }
\label{app_data}

\subsection{ Module Classes }

\begin{figure}[!htb]
\begin{center}
\includegraphics[width=\textwidth]{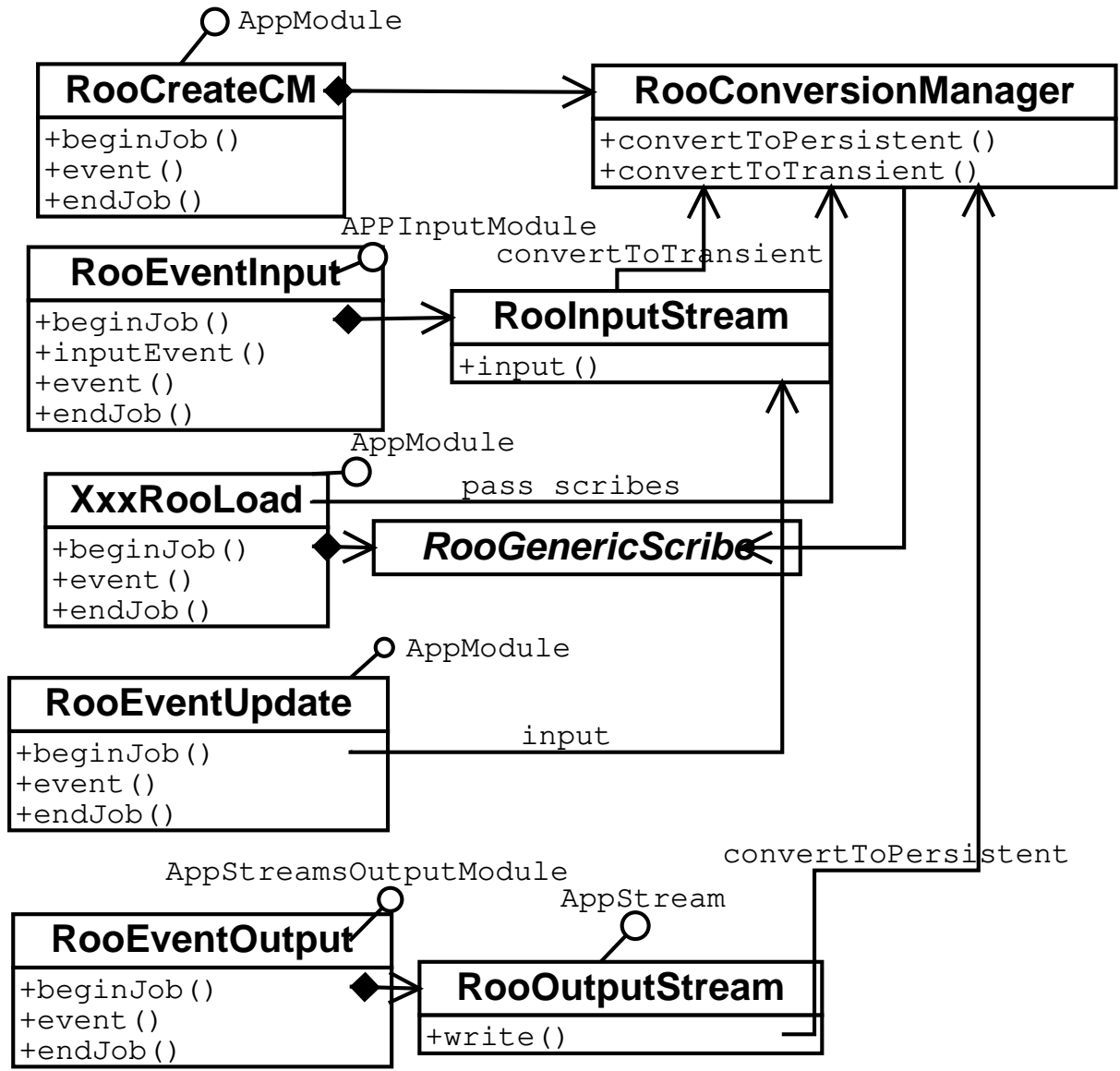}
\caption[ bla ]{ The figure shows an UML class diagram of the
framework modules and related classes of the \kanga\ system.  }
\label{fig_modules}
\end{center}
\end{figure}

Several module classes and other classes are used to implement the
ROOT based event data store in the framework. Figure~\ref{fig_modules}
shows a UML~\cite{uml} class diagram. 

\begin{description}

\item[RooOutputModule] This class inherits from a framework output
module base class, which uses framework output stream class instances
to output persistent objects to the \kanga\ data store for every event.
It is configured at run-time to set up output streams.

\item[RooOutputStream] This class inherits from and implements the
framework output stream class. Instances of this class control a
single tree in a ROOT file. They can open a new ROOT file, create a
tree and perform output of persistent objects to the tree using a
RooConversionManager class instance via the so-called {\em scribes}
connected to this stream.

\item[RooCreateCM] This module creates a new RooConversionManager
object for every event.

\item[XxxRooLoad] There are several modules which create 
scribes for each data type, e.g. AOD, TAG or micro-truth data.  These
modules create the correct scribes and can be configured with a
particular RooOutputStream or RooInputStream.  Each scribe is
associated with the RooOutputStream or RooInputStream and passed to
the RooConversionManager object.

\item[RooInputStream] This class inherits from and implements a
framework input stream class. It can open existing ROOT files and
perform input of persistent objects from a tree using the
RooConversionManager and the scribes connected to this stream.

\item[RooInputModule] This class inherits from a framework input
module class. It is configured at run-time to set up input streams.

\item[RooEventUpdate] This module uses RooInputStream class
instances to read persistent objects from trees for every
event using the RooConversionManager. 

\end{description}

\subsection{ Scribe Classes }

\begin{figure}[!htb]
\begin{center}
\includegraphics[width=\textwidth]{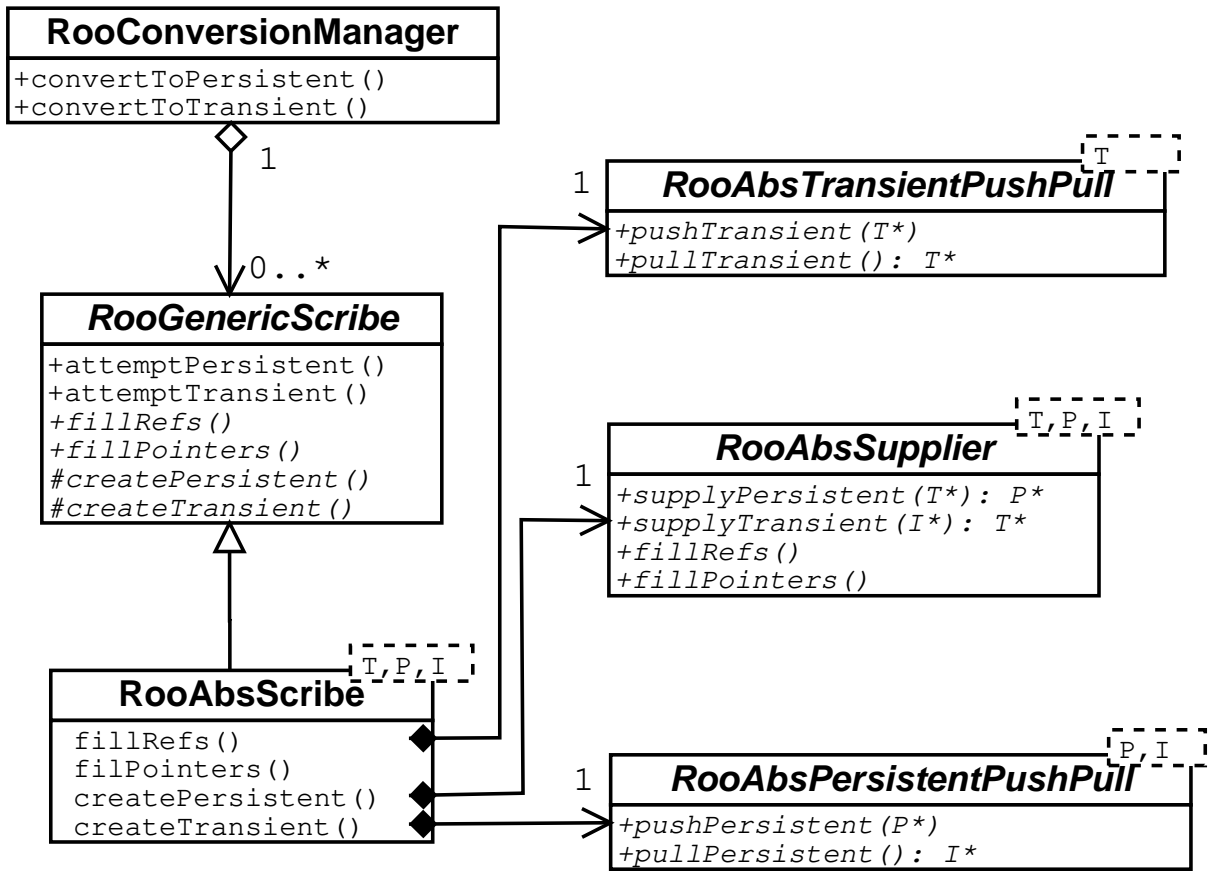}
\caption[ bla ]{ The figure shows an UML class diagram of the scribe
classes and their associated helper classes.  }
\label{fig_scribes}
\end{center}
\end{figure}

\begin{figure}[!htb]
\begin{center}
\includegraphics[width=\textwidth]{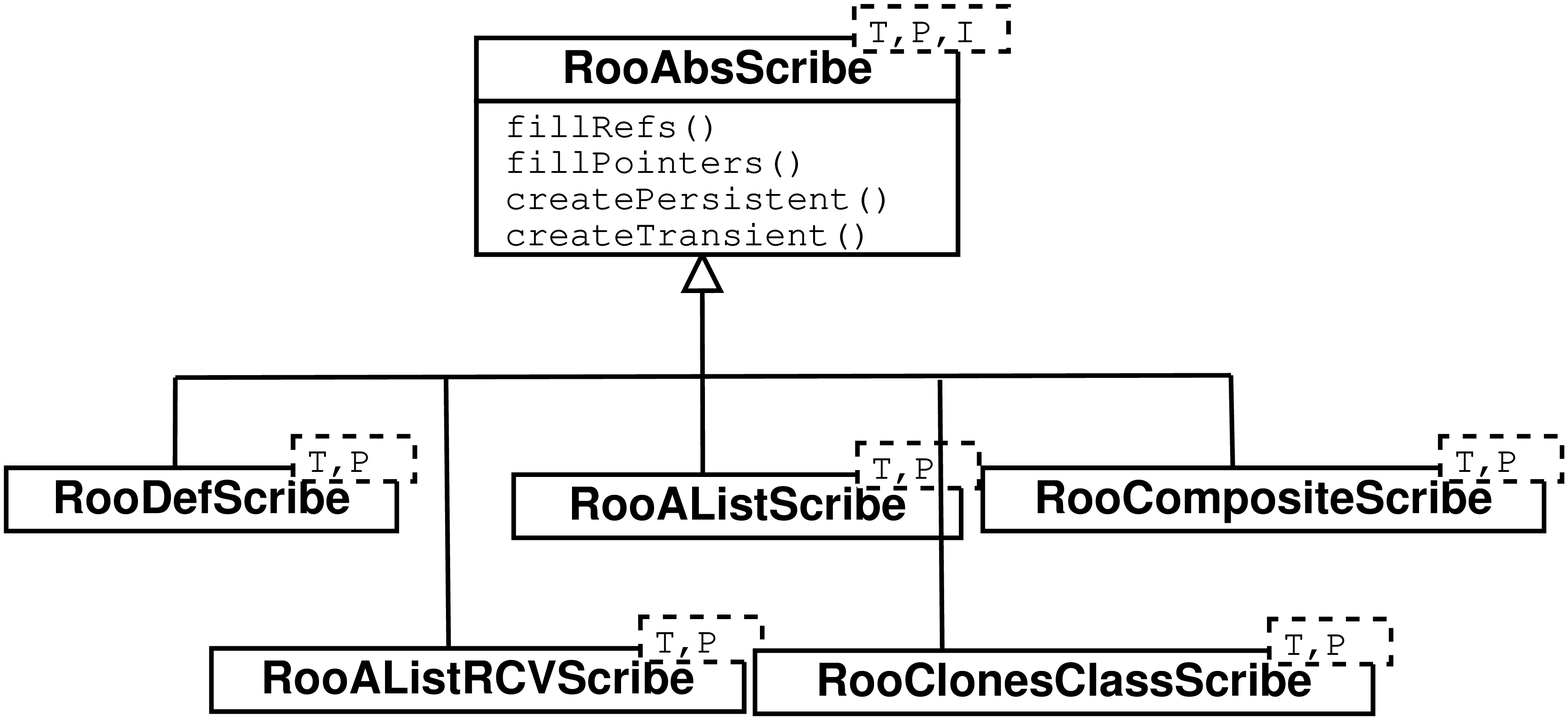}
\caption[ bla ]{ The figure shows the inheritance relationships
between various scribe classes as an UML class diagram. }
\label{fig_absscribe}
\end{center}
\end{figure}

A set of classes called scribes deals with the actual output and input
operations together with the RooConversionManager class. There are
scribes for one-to-one, many-to-many and many-to-one mappings of
transient-to-persistent objects. Figures~\ref{fig_scribes}
and~\ref{fig_absscribe} present UML class diagrams. 

\begin{description}

\item[RooConversionManager] This class uses lists of scribes it
receives to write or read event data. 

\item[RooGenericScribe] This class defines an interface to all scribe
objects to be used by the RooConversionManager and the XxxRooLoad
module classes. It provides generic methods to convert
transient to persistent objects and vice versa which must be
implemented by subclasses. 

\item[RooAbsScribe\btmpl{T,P,I}] This class implements the
RooGenericScribe and delegates to other classes the tasks of obtaining
(storing) transient objects T from (in) the event, obtaining (storing)
persistent objects P from (in) the \kanga\ data store and creating new
transient or persistent objects T or P. The class is templated by the
transient class T, persistent class P and interface class I to the
persistent class to obtain explicit type safety. The interface class I
is generally RooEvtObj\tmpl{T}.

\item[RooDefScribe\btmpl{T,P}] This class creates fully functional
scribe objects to convert single transient objects T to single
persistent objects P in the \kanga\ data store (one-to-one mapping) and
vice versa.

\item[RooAListScribe\btmpl{T,P}] This class is a scribe to convert 
collections of transient objects T (HepAList\tmpl{T} from
CLHEP~\cite{clhep}) to collections of persistent objects P in the ROOT
data store using the ROOT container class TObjArray (many-to-many
mapping). 

\item[RooAListRCVScribe\btmpl{T,P}] This scribe class provides
conversion of collections of static transient objects T
(HepAList\tmpl{T}) to collections of static persistent objects P in
the \kanga\ data store using the container class RooClonesVectorR\tmpl{P}
and vice versa (many-to-many mapping).

\item[RooCompositeScribe\btmpl{T,P}] This scribe class implements a 
conversion of collections of transient objects T (HepAList\tmpl{T}) to
a composite persistent object P (many-to-one mapping).

\item[RooClonesClassScribe\btmpl{T,P}] This scribe class implements a 
conversion of collections of static transient objects T
(HepAList\tmpl{T}) to a composite persistent object P which uses the
ROOT class TClonesArray with reuse of memory for new persistent
objects (many-to-one mapping).

\end{description}

\subsection{ Classes to handle Transient and Persistent Objects }

\begin{figure}[!htb]
\begin{center}
\includegraphics[width=0.5\textwidth]{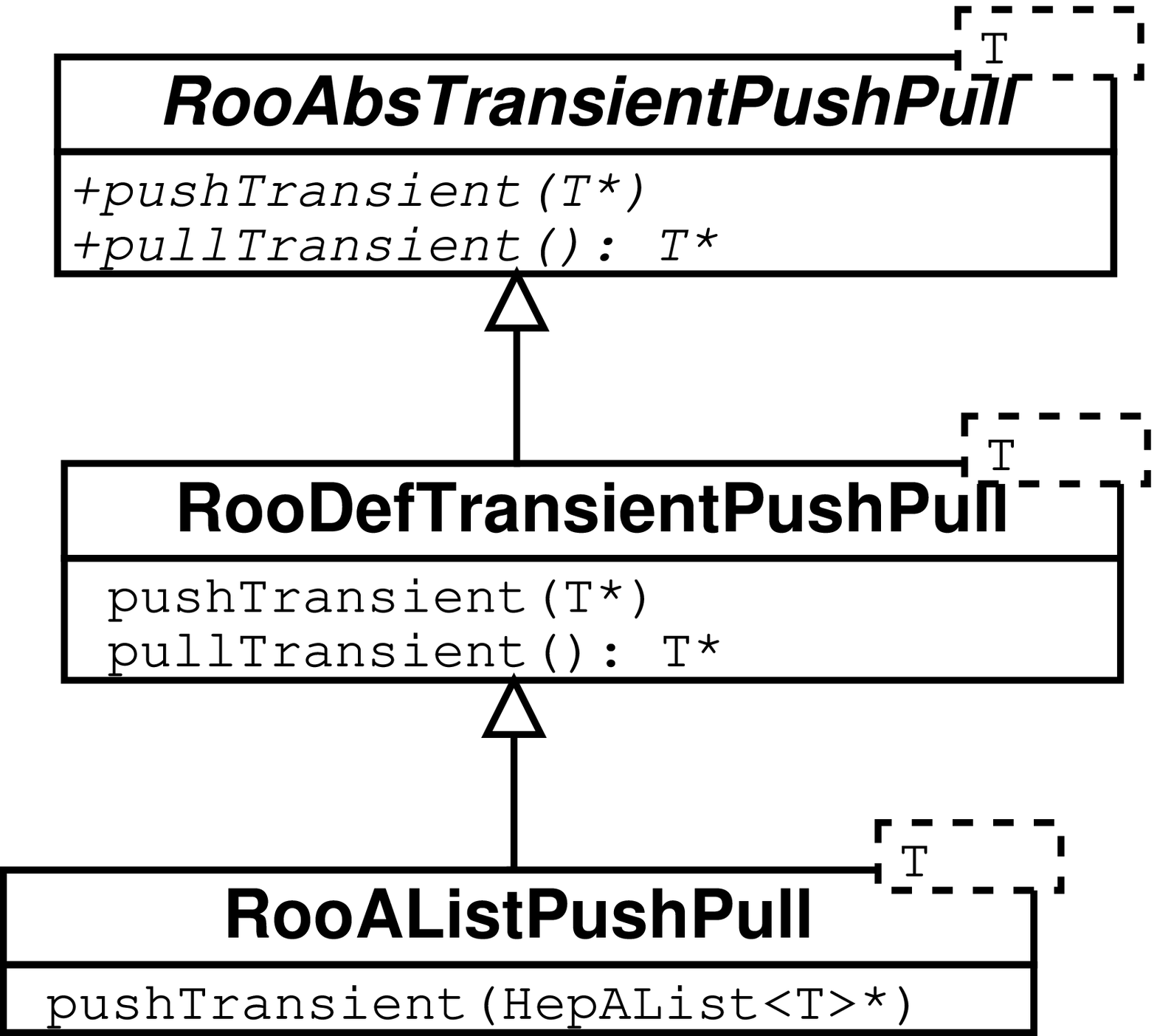}
\caption[ bla ]{ The figure shows the inheritance relationships
between the transient pushpull classes as an UML class diagram.  }
\label{fig_transpushpull}
\end{center}
\end{figure}

\begin{figure}[!htb]
\begin{center}
\includegraphics[width=\textwidth]{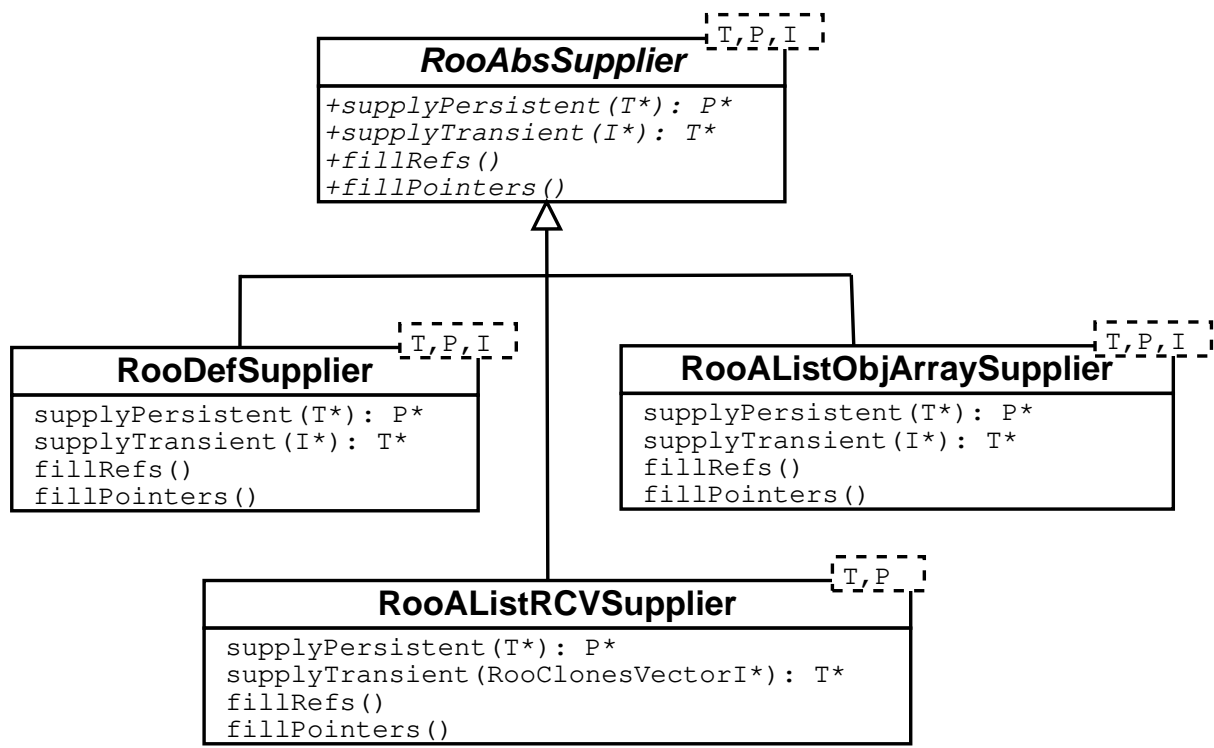}
\caption[ bla ]{ The figure shows the inheritance relationships
between the supplier classes as an UML class diagram. }
\label{fig_supplier}
\end{center}
\end{figure}

\begin{figure}[!htb]
\begin{center}
\includegraphics[width=0.5\textwidth]{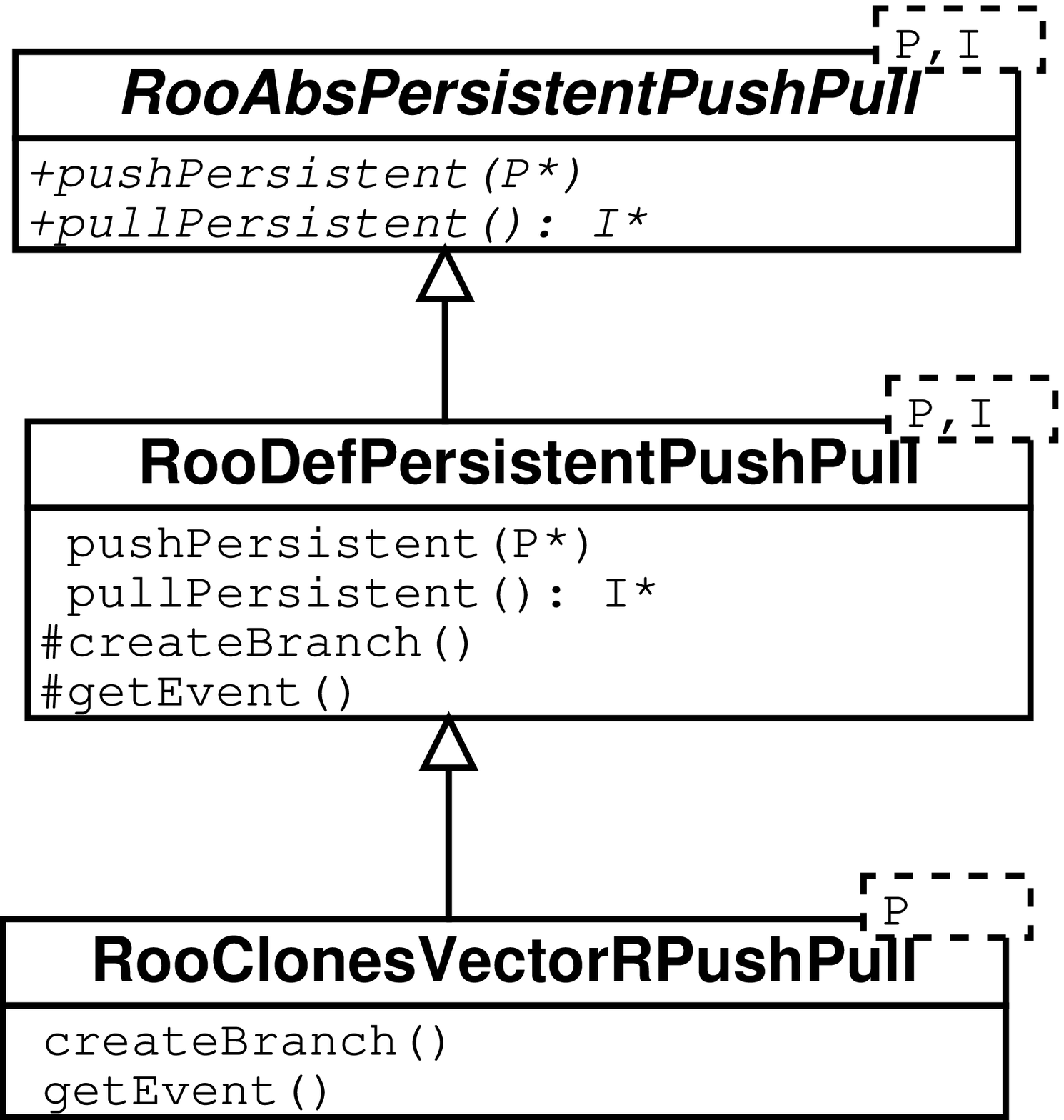}
\caption[ bla ]{ The figure shows the inheritance relationships
between the persistent pushpull classes as an UML class diagram. }
\label{fig_perspushpull}
\end{center}
\end{figure}

The tasks delegated by RooAbsScribe\tmpl{T,P,I} are implemented by the
so-called {\em pushpull} and {\em supplier}
classes. Figures~\ref{fig_scribes}, \ref{fig_transpushpull},
\ref{fig_supplier} and \ref{fig_perspushpull} show UML class
diagrams. 

\begin{description}

\item[RooAbsTransientPushPull\btmpl{T}] This class sets up an
interface for storing (obtaining) transient objects T in (from) the
event. It declares a method for storing (pushing) and a method for
obtaining (pulling) transient objects.

\item[RooAbsPersistentPushPull\btmpl{P,I}] This class declares the
interface for writing (reading) persistent objects P to (from) the
\kanga\ data store using one method for writing (pushing) and one for
reading (pulling). Reading from the \kanga\ data store is performed
through the interface class I of the persistent class P.

\item[RooTransientPushPull\btmpl{T}] This class inherits from
RooAbsTransientPushPull\tmpl{T} and adds the necessary functionality 
so that subclasses can obtain transient objects T from the event or
put transient objects T into the event.

\item[RooPersistentPushPull\btmpl{P,I}] This class is derived from
RooAbsPersistentPush-Pull\tmpl{P,I} and adds the necessary
functionality so that subclasses can operate branches in the tree
and read persistent objects P from the \kanga\ data store or write
persistent objects P to the \kanga\ data store. 

\item[RooDefTransPushPull\btmpl{T}] This class implements the
RooTransientPushPull class to provide input and output of single
transient objects via the interface dictionary access functions of the
event. 

\item[RooAListPushPull\btmpl{T}] This class is a subclass of
RooDefTransPushPull\tmpl{T} and provides a specialised method to
store a HepAList\tmpl{T} object in the transient event. 

\item[RooDefPersPushPull\btmpl{P,I}] This class inherits from
RooPersistentPushPull\tmpl{P,I} and provides implementations for the 
methods declared in RooAbsPersistentPush-Pull\tmpl{P,I} to read
persistent objects P through the interface I from the ROOT event store
or write persistent objects P to the \kanga\ data store. 

\item[RooClonesVectorRPushPull\btmpl{P}] This class inherits from
RooDefPersPushPull\tmpl{P,I} and provides special implementations of
the methods to create new branches and to read persistent objects P
from the \kanga\ data store when the special container class
RooClonesVectorR\tmpl{P} is used.

\item[RooAbsSupplier\btmpl{T,P,I}] This class declares the interface
to create a persistent object P using a transient object T for writing
to the \kanga\ data store and a method to create a transient object T
from a persistent object P through its interface I.

\item[RooDefSupplier\btmpl{T,P,I}] This class implements the
RooAbsSupplier\tmpl{T,P,I} class for transient objects T
and persistent objects P where the persistent class P is a subclass of
the interface I (usually I=RooEvtObj\tmpl{T}). 

\item[RooAListObjArraySupplier\btmpl{T,P,I}] This class implements 
the many-to-many mapping of collections of transient objects T in
HepAList\tmpl{T} and the ROOT class TObjArray. It
is a subclass of RooAbs-Supplier\tmpl{T,P,I}.

\item[RooAListRCVSupplier\btmpl{T,P}] This class implements
RooAbsSupplier\tmpl{T,P,I} for the many-to-many mapping of collections
of static transient objects T in HepA-List\tmpl{T} and
RooClonesVectorR\tmpl{P}. The interface to RooClonesVectorR\tmpl{P} is
RooClonesVectorI (derived directly from RooPersObj) and thus not
needed as a template parameter.

\end{description}

\subsection{ Persistent Classes }
\label{sec_pers}

Persistent classes are defined using the following base class
hierarchy. All classes in this list are dressed with special ROOT
macros for full integration into the ROOT system.
Figure~\ref{fig_persistent} presents a UML class diagram.

\begin{description}

\item[RooPersObj] This class provides the connection with the ROOT
system by inheriting from the ROOT class TObject and creates an object
identifier for each persistent class which is unique within one
session. 

\item[RooEvtObj\btmpl{T}] This class is a subclass of RooPersObj and
declares the interface to all persistent classes. It declares a method
to create a new transient object T which the concrete persistent class
must implement.

\item[XxxDataR] A concrete persistent class for storage of the data
contained in a transient class T in a package Xxx, e.g. the TAG data
or the AOD data, must inherit from RooEvtObj\tmpl{T}. It implements
the method to create a new transient object T from its data and must
provide a constructor which creates a new persistent object using a
transient object T.

\item[RooClonesVectorI] This is the base class for all
RooClonesVectorR\tmpl{P} classes. It inherits directly from RooPersObj
and handles memory management for the TClonesArray objects contained
in RooClonesVectorR\tmpl{P} together with
RooClones-VectorRPushPull\tmpl{P}. The pushpull object fetches the
pointer to the TClonesArray created by the constructor of
RooClonesVectorR\tmpl{P} when the first event is read in and passes
that pointer to a static data member of this class.

\item[RooClonesVectorR\btmpl{P}] This container class for static
persistent objects P uses the ROOT class TClonesArray internally which
supports reuse of memory for new persistent objects using the C++
construct {\em new with placement}\footnote{ The C++
new-with-placement allows to create a new object using the already
allocated memory of an existing object}. In this way the time
consuming operations of allocation and release of memory for the
persistent objects of each event are minimised. This class's
constructor uses a TClonesArray from the base class static data member
when present instead of creating a new TClonesArray.

\end{description}

\bibliographystyle{unsrt}
\bibliography{kanga}

\clearpage

\section*{ Figures }

% The End
\end{document}